\documentclass[asp,rsi,amsmath,amssymb,preprint,nofootinbib]{revtex4-1}

\usepackage{graphicx}
\usepackage{dcolumn}
\usepackage{bm}
\usepackage{amsmath}

\begin{document}


\title{F-term Inflation and $RSII$ Brane Model}

\author{M. Roostaee}
\email{m\_roustaie62@ut.ac.ir}

\author{Amir M. Abbassi}
\email{amabasi@khayam.ut.ac.ir}
\affiliation{Department of Physics, University of Tehran,\\P.O. Box 14395/547, Tehran, Iran.}
\date{\today}

\begin{abstract}
Considering hybrid F-term inflation in $RSII$ model and using the most recent data from \textit{Planck 2018} and in comparison with \textit{Planck 2015} \textit{WMAP 9-year} , we can obtain some interesting constraints on main parameters of the model. Also, we attain convenient compatibility between this model and observational data. We show that this setup provides a successful hybrid inflation with high enough Reheating Temperature, $T_R$, to have a successful thermalization.\\
After inflation, particles are created in the process of preheating. Inflaton field(s) oscillated and fermionic field(s) interacted with it in a non-perturbative regime of parametric resonance. We apply theory of fermionic preheating coupling to the inflaton, without expansion of the universe, to calculate the occupation number of created particles analytically, and some interesting results are achieved.
\end{abstract}


\maketitle


\section{Introduction}
Over the last years there has been considerable interest in higher dimensional cosmological models. Because of high consistency with observational results, evolution of the universe in most models is described by supersymmetric hybrid inflation \cite{1},\cite{2}. Hybrid inflation was introduced to solve some problems in inflationary models \cite{3}, \cite{4} for example; F-term hybrid inflation model overcome blue spectrum problem \cite{5},\cite{6}. Thus, it is very considerable to look for a generalized supersymmetric braneworld inflation consistent with recent observations\cite{7}. However inflationary scenarios overcome some cosmological problems, but exiting of inflation and entering matter dominated era after a suitable reheating process is still challenging. At the other hand some elementary particles can make shortcut to attain this purpose; i.e. sterile neutrinos \cite{8}. Searching for sterile neutrinos is still in process by searching cosmological signatures \cite{9} and providing high-energy collides in CERN \cite{10} and other experiments \cite{11}, \cite{12}.\\
In Randall-Sundrum braneworld scenario\cite{14}, we suppose that our 4D Universe is living on a 3-brane, embedded in a 5D bulk with an extra space dimension. We add some supersymmetric hybrid inflation to this context and hope to obtain some interesting cosmological implications.\\
In the present paper,we are interested on F-term effect on some inflationary aspects like observable quantities and reheating temperature. We briefly reviewed the setup of F-term potential and loop corrections on it \cite{15},\cite{16} and \cite{17}. In this model we note that the F-term will dominate. This case was considered for the first time in \cite{18}.\\
In the following sections,after a brief review of F-term inflation; we will recall some foundation of field equations on the brane and also basis of inflation dynamics. Then we consider the extension of this model on $RS II$ brane and calculate some perturbative parameters in this setup and the results are in good agreement with  recent $ \textit{WMAP 9-year} $ and Planck 2015 and Planck 2018 observations \cite{19}, \cite{7}, \cite{PlanckVI}, \cite{PlanckX}. Then we briefly study reheating after inflation and some interesting results are obtained.
In the last section we develop the theory of fermionic preheating for non-expanding universe on the brane.
\section{F-term Model}
The most interesting SUSY hybrid inflation models are F-term inflation and D-term inflation. F-term attracts more attention than the other because it is tailor made to fit with Higgs' mechanism. Here we review the structure of F-term inflation which is a special case for P-term inflation. By adding Fayet-Illiopoulos terms $ \vec \xi $ to the theory , as mentioned in \cite{17} , $ SU(2,2|2) $ symmetry breaks down to $ N=2 $ supersymmetry.
\begin{eqnarray}
\xi  \equiv \sqrt {|\vec \xi |^2}  = \sqrt {\xi _ + \xi _ -  + (\xi _3)^2 }&,&\xi _ \pm  \equiv \xi _1 \pm i\xi _2 .
\label{eq1}
\end{eqnarray}
With considering global supersymmetry, P-term inflation is derived from the superpotential
\begin{equation}
W = \sqrt 2 gS\left( {{\phi _ + }{\phi _ - } - \frac{{{\xi _ + }}}{2}} \right) ,
\label{eq2}
\end{equation}
where $ S $, $ {\phi _ + } $, $ {\phi _ - } $ are chiral superfields with positive coupling constant g and charges $ {Q_{S}} =  0 $, $ Q_{\phi _ + } =  + 1 $, $ Q_{\phi _ - } =  - 1 $ respectively. with this superpotential we can obtain scalar potential
\begin{eqnarray}
V = |\partial W|^2 +{ \frac{g^2}{2}}{D^2}&,&D = |\phi _ + |^2 - |\phi _ - |^2 - \xi _3 .
\label{eq3}
\end{eqnarray}
If we choose $ \xi _ +  = \xi _ -  = \xi  = 2M^2 > 0 $ and $ \xi _3 = 0 $ we recover the potential of F-term inflation model with $ W = \sqrt 2 gS\left( {\phi '}_ + {\phi '}_ -  - M^2 \right) $ and $ D = |{\phi '}_ + |^2 - |{\phi '}_ - |^2 $
\begin{equation}
\begin{split}
V_{N = 2}^F &= 2{g^2}\left( {|S{{\phi '}_ + }{|^2} + |S{{\phi '}_ - }{|^2} + |{{\phi '}_ + }\phi ' - {M^2}{|^2}} \right) 
\\ &
+ \frac{{{g^2}}}{2}\left( {|{\phi _ + }{|^2} - |{\phi _ - }{|^2}} \right) .
\end{split}
\label{eq4}
\end{equation}
Now we need to determine the vacua of P-term model which are :\\
\textbf{1. Local minimum} with flat direction of scalar field S which correspond to a \textit{de-Sitter minima} or de-Sitter solution, and S provides a flat direction in potential :
\begin{eqnarray}
\phi _ + = \phi _ -  = 0,&V_0 = \frac{1}{2}{g^2 \xi ^2},&|S|^2 > S_c^2 \equiv \frac{\xi }{2} .
\label{eq5}
\end{eqnarray}
\textbf{2. Global minimum} with choosing suitable components has a solution ,
\begin{equation}
\left. {\begin{array}{*{20}{c}}
{|\phi _ + |^2 - |\phi _ - |^2 = \xi _3}\\
{|\phi _ + |^2 + |\phi _ - |^2 = \xi }
\end{array}} \right\} \Rightarrow \begin{array}{*{20}{c}}
{|\phi _ - |^2 = \frac{\xi  - \xi _3}{2}}\\
{|\phi _ + |^2 = \frac{\xi  + \xi _3}{2}}
\end{array}
\label{eq6}
\end{equation}
which for F-term model with $ \xi _3 = 0 $ reduces to $ |\phi _ - |^2 = |\phi _ + |^2 = \frac{\xi }{2} $ .
\subsection{Gauge Theory Loop Corrections }
Due to the first loop corrections in gauge theory, the flat direction of the inflation field S is uplifted. Using \textit{Coleman-Weinberg} formula \cite{20}  one can find the effective 1-loop potential for large inflaton field $ S $ 
\begin{equation}
V_{1 - loop} = \frac{g^2 \xi ^2}{2}\left( 1 + \frac{g^2}{8\pi ^2}\ln \frac{|\phi |^2}{|\phi _c|^2} \right) .
\label{eq7}
\end{equation}
This term is important because leads to the motion of the field $ S $ towards the bifurcation point and the end of inflation.\\
For F-term model all non-gravitational higher loop corrections are finite. The radiative corrections obtained above lead to the effective potential for f-term inflation
\begin{equation}
V^F(\phi ) = \frac{g^2 \xi ^2}{2}\left( {1 + \frac{g^2}{8\pi ^2}\ln \frac{|\phi |^2}{|\phi _c|^2} + \frac{|\phi |^4}{8} +  \cdots } \right) .
\label{eq8}
\end{equation}
During inflation potential values for both D-term and F-term model are practically equal; but different values of $ V^\prime $ make perturbations a little smaller in F-term model.
\section{Field Equations on the Brane }
Now we review the basic equations on the brane \cite{21} , \cite{23} , \cite{24}. With supposing a Randall-Sundrum II model, a 5D cosmological constant in the bulk, $ \Lambda _5 $ , matter contribution on the brane with energy-momentum tensor $ \tau _{mu \nu } $ , tension of the brane $ \lambda $ and 5D Planck mass $ M_5 $ we have :
‎\begin{subequations}
\begin{equation}
G_{\mu \nu } + \Lambda _4 g_{\mu \nu } = \frac{8\pi }{M_p^2}{\tau _{\mu \nu }} + \left( {\frac{8\pi }{M_5^3}} \right)^2 \pi _{\mu \nu } - E_{\mu \nu } ,
\label{eq9}
\end{equation}
\begin{equation}
\pi _{\mu \nu } = \frac{1}{12}\tau \tau _{\mu \nu } + \frac{1}{8}g_{\mu \nu } \tau _{\alpha \beta } \tau ^{\alpha \beta } - \frac{1}{4} \tau _{\alpha \mu } \tau _\mu ^\alpha  - \frac{1}{24} \tau ^2 g_{\mu \nu } ,
\label{eq10}
\end{equation}
\begin{equation}
E_{\mu \nu } \equiv C_{\beta \rho \sigma }^\alpha {n_\alpha }{n^\rho } g_\mu ^\beta g_\nu ^\sigma  ,
\label{eq11}
\end{equation}
\end{subequations}
where the $ g_{\mu \nu } $ is the induced 4D metric, $ \Lambda _4 $ the effective 4D cosmological constant, $ M_p $ the usual 4D Planck mass, $ n^\alpha $ normal vector on the brane and $ E_{\mu \nu } $ electric part of the Weyl tensor. Which
\begin{eqnarray}
M_p = \sqrt {\frac{3}{4\pi }} \frac{M_5^3}{\sqrt \lambda  }&,&\Lambda _4 = \frac{4\pi }{M_5^3}\left( {\Lambda _5 + \frac{4\pi \lambda ^2}{3M_5^3}} \right) .
\label{eq12}
\end{eqnarray}
In a cosmological model, which induced metric $ g_{\mu \nu } $ on the brane has the form of spatially flat FRW form and $ a(t) $ is the scale factor, then the Friedmann-like equation on the brane takes the generalized form \cite{25}, \cite{26}
\begin{equation}
H^2 = \frac{\Lambda _4}{3} + \frac{8\pi }{3M_p^2}\rho  + \left( {\frac{4\pi }{3M_5^3}} \right)^2 \rho ^2 + \frac{c}{a^4} ,
\label{eq13}
\end{equation}
where $ C $ is an integration constant term arising from $ E_{\mu \nu } $ and reminds radiation term or "Dark Radiation" term. This term disappears quickly during inflation and can be ignored. On the critical brane, $ \Lambda _4 = 0 $ , we have
\begin{equation}
\Lambda _5 =  - \frac{4\pi \lambda ^2}{3M_5^3} .
\label{eq14}
\end{equation}
So the Friedmann generalized equation has the form
\begin{equation}
H^2 = \frac{8\pi }{3M_p^2}\rho \left( {1 + \frac{\rho }{2\lambda }} \right) .
\label{eq15}
\end{equation}
In the low energy regime, $ \rho  \ll \lambda $ , we recover usual Friedmann equation. But in the high energy regime, $ \rho  \gg \lambda $ , the generalized Friedmann equation takes the form
\begin{equation}	
H^2 = \frac{4\pi \rho ^2}{3\lambda M_p^2} .
\label{eq16}
\end{equation}
We will use this regime for some estimations in the following sections.
\section{Inflation on the $RSII$ Braneworld model}
L. Randall and R. Sundrum constructed two 5D universe models to overcome hierarchy problem and the weakness of gravity in \cite{13} and \cite{14}, but the second one attracts more interest and has a very good agreement with observations. In this cosmological scenario, as already mentioned, we will consider a scalar field as inflaton field. We confine scalar field $ \phi $ on the brane with a self-interacting potential $ V(\phi ) $ given in \cite{27}. Scalar field $ \phi $ obeys the homogenity and isotropy of the 4D universe, then $ \phi $ is a function of time only. This homogeneous field $ \phi (t) $ behaves like a perfect fluid with energy density $ \rho (t) = \frac{1}{2}{\dot \phi }^2 + V(\phi ) $ and pressure $ p(t) = \frac{1}{2}{\dot \phi }^2 - V(\phi ) $ . There is no energy flow between the brane and the bulk, so the energy-momentum tensor $ T_{\mu \nu } $ of $ \phi $ is conserved, which means $ \nabla ^\nu T_{\mu \nu } = 0 $ . Then we obtain the continuity equation for the $ \rho $ and $ p $
\begin{equation}
\dot \rho  + 3H(p + \rho ) = 0 ,
\label{eq17}
\end{equation}
where H is the Hubble parameter $ H = \frac{\dot a}{a} $ .\\
Finally we find the equation of motion for the scalar field $ \phi $
\begin{equation}
\ddot \phi  + 3H\dot \phi  + V'(\phi ) = 0 .
\label{eq18}
\end{equation}
Inflation occurs in early universe when matter density is so high, then Friedmann equation takes the form
\begin{equation}
H^2 = \frac{4\pi \rho ^2}{3\lambda M_p^2} .
\label{eq19}
\end{equation}
In the slow-roll approximation, there is two constraints with slow-roll parameters $ \epsilon $ and $ \eta $ which must be satisfied
\begin{eqnarray}
\epsilon \equiv  - \frac{\dot H}{H^2},&\eta  \equiv \frac{V''}{3H^2} = \frac{\dot\epsilon }{H},& \epsilon\ll 1,|\eta | \ll 1 .
\label{eq20}
\end{eqnarray}
In this approximation the equation of motion of the scalar field takes the form
\begin{equation}
\begin{array}{*{20}{c}}
{\begin{array}{*{20}{c}}
{|\ddot \phi | \ll 3H\dot \phi }\\
{|\ddot \phi | \ll V'}
\end{array}}& \Rightarrow &{\dot \phi  \simeq  - \frac{V'}{3H}}
\end{array} ,
\label{eq21}
\end{equation}
and the generalized Friedmann equation becomes
\begin{equation}
V \gg {\dot \phi }^2 \Rightarrow \rho  \approx V(\phi ) \Rightarrow H^2 \simeq \frac{4\pi V^2}{3\lambda M_p^2} .
\label{eq22}
\end{equation}
In this regime with slow-roll approximation, $ \rho  \gg \lambda  \to V \gg \lambda $, the slow-roll parameters and the number of e-folds are given by
\begin{subequations}
\begin{equation}
\epsilon \simeq \frac{M_p^2}{16\pi } \left( \frac{V'}{V} \right)^2 \frac{\left( 1 + \frac{V}{\lambda } \right)} {\left( 1 + \frac{V}{2\lambda } \right)^2} ,
\label{eq23}
\end{equation}
\begin{equation}
\eta  \simeq \frac{M_p^2}{8\pi }\left( {\frac{V''}{V}} \right) \frac{1}{\left( {1 + \frac{V}{2\lambda }} \right)} ,
\label{eq24}
\end{equation}
\begin{equation}
N \simeq  - \frac{8\pi }{M_p^2} \int_{\phi _i}^{\phi _f} {\frac{V}{V'}\left( {1 + \frac{V}{2\lambda }} \right) d\phi } ,
\label{eq25}
\end{equation}
\end{subequations}
where $ V'' = \frac{d^2 V}{d\phi ^2} $ and the subscripts $ i $ and $ f $ are used to denote the epoch when the cosmological scales exit the horizon and the end of inflation, respectively. $ \epsilon  \ll 1 $ and $ |\eta | \ll 1 $ warrants inflation and inflation ends at $ \phi_f = \phi_{end} $ which makes $ \max (\epsilon ,|\eta |) = 1 $ . For a strong enough inflation we take $ 50 \le N \le 60 $. Another important cosmological constraint comes from the power spectrum of the curvature perturbations \cite{21}
\begin{equation}
P_R(k) \simeq \frac{128\pi }{3M_p^6}\frac{V^3}{{V'}^2}\left( {1 + \frac{V}{2\lambda }} \right)^3 .
\label{eq26}
\end{equation}
The amplitude of tensor perturbations defined by \cite{30}
\begin{equation}
P_g(k) \simeq \frac{128}{3M_p^4} V \left( {1 + \frac{V}{2\lambda }} \right)^2 F^2(x) ,
\label{eq27}
\end{equation}
where 
\[ x=H M_p \sqrt {\frac{3}{4\pi \lambda }} \]
and
\[ F^2(x)=\left[ \sqrt {1 + x^2}  + x^2 \ln \left( \frac{1}{x} + \sqrt {1 + \frac{1}{x^2}} \right) \right]^{ - 1}. \]
The ratio of tensor to scalar perturbations and the running of the scalar index respectively are
\begin{equation}
r(k) \simeq \left( {\frac{M_P^2}{\pi } \frac{{V'}^2}{V^2} \frac{F^2(x)}{\left( {1 + \frac{V}{2\lambda }} \right)^2}} \right)_{k = {k_*}} ,
\label{eq28}
\end{equation}
\begin{equation}
\frac{dn_s}{d\ln k} \simeq \frac{M_P^2}{4\pi }\frac{V'}{V}\frac{1}{\left( {1 + \frac{V}{2\lambda }} \right)} \left( {3\frac{\partial \epsilon }{\partial \phi } - \frac{\partial \eta }{\partial \phi }} \right) ,
\label{eq29}
\end{equation}
where the right-hand side evaluated at the horizon-crossing when the co-moving scale equals the Hubble radius during inflation and $ k_*=Ha$ . The spectral index for the scalar perturbations is given in terms of the slow-roll parameters
\begin{equation}
n_s - 1 \equiv \frac{d\ln P_R}{d\ln k} = 2\eta  - 6\epsilon .
\label{eq30}
\end{equation}
\section{\label{sec5}F-term Inflation and Planck 2018 results}

In this section, we introduce a limit on the brane tension, $ \lambda $. With this goal in mind, we investigate the F-term inflation. Slow-roll parameter $ \eta $ determines the end of inflation which ends at $ \phi  = \phi _c $ that $ \phi _c \ge \phi _{end} $. In the frame-work of RSII model, one can easily obtain \cite{29}
\begin{eqnarray}
|\eta | = 1& \Rightarrow &\phi _{end} =\frac{M_p g^2 \xi }{8 \pi ^{3/2}} \sqrt {\frac{2}{g^2 \xi ^2} \left( \frac{g^2 \xi ^2}{4\lambda} + 1 \right)} ,
\label{eq31}
\end{eqnarray}
which enables us to find an upper limit for the brane tension $ \lambda $
\begin{eqnarray}
\phi _c \ge \phi _{end} & \Rightarrow & \lambda  \le \frac{g^2 \xi ^2}{\frac{M_P^2 g^2}{8 \pi ^3 \xi } - 4} .
\label{eq32}
\end{eqnarray} 
Considering eq.(\ref{eq25}), we can evaluate $ \phi_* $ (the inflaton field at horizon crossing) as
\begin{equation}
\phi _*^2 \simeq \frac{N g^2M_P^2}{16\pi ^3 \left( 1 + \frac{g^2 \xi ^2}{4\lambda } \right)}+\xi .
\label{eq33}
\end{equation}
The combination of $ \textit{WMAP 9-year} $ and $ Planck 2015 $ data gives the following results \cite{19}, \cite{7}

\begin{subequations}
‎\begin{align}‎
n_s & = 0.9603 \pm 0.0073 ,\\
r & < 0.099 ,\\
\frac{{d{n_s}}}{{d\ln k}} & = -0.0134 \pm  - 0.0090 ,\\
P_R(k) & = (2.23 \pm 0.16) \times {10^{ - 9}} ,\\
\epsilon _V & < 0.008 ,\\
\eta _V & =  - 0.010_{ - 0.011}^{ + 0.005} .
\label{eq34}
\end{align}
\end{subequations}

On the other hand the combination of Planck 2018  \footnote{TT,TE,EE+lowE+lensing} and $ BAO $ and $ BK14 $ data gives the following results \cite{PlanckVI}, \cite{PlanckX}

\begin{subequations}
‎\begin{align}‎
n_s & = 0.9649 \pm 0.0067 ,\\
r & < 0.076 ,\\
\frac{{d{n_s}}}{{d\ln k}} & = -0.0045 \pm  - 0.0067 ,\\
P_R(k) & = (2.22 \pm 0.16) \times {10^{ - 9}} ,\\
\epsilon _V & < 0.0097 ,\\
\eta _V & =  - 0.010_{ - 0.011}^{ + 0.007} .
\end{align}
\label{eq34-1}
\end{subequations}
which is consistent with $ Gaia $ results \cite{Riess}.
\\
Considering eq.(\ref{eq30}) , $ g > \sqrt {\frac{32 \pi ^3\xi }{M_P^2}} $ with $ \xi  = 32 \times 10^{ - 8} M_P^2 $ and $ M_P = 1.2 \times 10^{19} GeV $ \cite{22} , we find a limit on coupling constant $ \sqrt 2  g > 0.02 \;  $.\\
As an example, for $ g=0.02 $ and $ N=60 $ from eq.(\ref{eq32}) we find an upper limit on brane tension :
\begin{equation}
\lambda  \le 3.941 \times 10^{-17} M_P^4 .
\label{eq35}
\end{equation} 
Evaluating $ \lambda  = 2.876 \times 10^{-17 }M_P^4 $ which satisfies the inequality (\ref{eq35}), we obtain:
\begin{equation}
P_R(k) \simeq 2.35 \times 10^{ - 9} ,
\label{eq36}
\end{equation}
which practically coincides with observational value.\\
As we mentioned before, large interval of values of $ N $ can reproduce the observation results as shown in the following table.
\begin{center}
\begin{tabular}{|c|c|c|c|}
\hline $ Inflation Parameters $ & $ n_s $ & $ r $ & $ dn_s/d \ln k $ \\ 
\hline $ N=50 $ & $ 0.9608 $ & $ 0.43*10^{-5} $ & $ -0.0014 $ \\ 
\hline $ N=55 $ & $ 0.9623 $ & $ 0.39*10^{-5} $ & $ -0.0012 $ \\ 
\hline $ N=60 $ & $ 0.9667 $ & $ 0.37*10^{-5} $ & $ -0.0011 $ \\ 
\hline 
\end{tabular} 
\end{center}
We can also remark that according to $ N $ the scalar spectral index $ n_s $ and the running of the spectral index $ dn_s/d \ln k $ increase, whereas, the ratio of the tensor to scalar perturbations $ r $ decreases.
\section{Reheating of the Universe}
During inflation Universe expands exponentially and cools down typically from temperature $ 10^{27} k $ to $ 10^{22} k $ \cite{18} (The exact drop of temperature is model dependent).\\
When inflation ends the temperature must return to the pre-inflationary temperature; in the aim of entering radiation dominated era \cite{30}. This is reheating or thermalization, which inflaton field decays into elementary particles and large amount of potential energy provides the high temperature we need for the Standard Hot Big Bang picture. At this stage Universe fills with standard model particles and the radiation dominated phase of the Universe begins. After that the temperature must provide a good basis for baryogenesis and nucleosynthesis. Baryogenesis step needs energies larger than electroweak scale and nucleosynthesis requires that the Universe be in a vicinity of thermal equilibrium at a temperature around $ {1 MeV} $.\\
The so-called "old" version of reheating was developed immediately after first inflationary scenarios. In old reheating theory, the inflaton field decays to scalar particles with finite probability of decaying. These decays can be formulated by coupling the inflaton field, $ \phi $, to other scalar ($ \chi $) or fermion ($ \psi $) fields through term in the lagrangian \cite{31} , \cite{5} such as $ \nu\sigma\phi\chi^2 $ and $ h \phi \bar \psi \psi $ 
\[ L = \frac{1}{2}{({\partial _\mu }\phi )^2} - \frac{{m_\phi ^2}}{2}{\phi ^2} - \frac{{m_\chi ^2}}{2}{\chi ^2} + \bar \psi \left( {i{\gamma _\mu }{\partial _\mu } - {m_\psi }} \right)\psi  \]
\begin{equation}
 + \nu \sigma \phi {\chi ^2} - h\bar \psi \psi \phi  - \Delta V(\phi ,\chi ) ,
\label{eq37}
\end{equation}
 which $ \sigma $ is a parameter with dimensionality of mass and $ \nu,h $ are dimensionless coupling constants and $ \Delta V $ includes terms with higher order than $ (\phi^2 , \chi^2) $. With dimensional analysis, since $ \Gamma  = t^{ - 1} = m $ one can easily obtain the tree-level decay rates $ (\Gamma) $ for these two interactions. When the mass of the inflaton is much larger than hose of $ \chi $ and $ \psi $ $ (m_{\phi} \gg m_{\chi} , m_{\psi}) $, the decay rate is 
 \begin{eqnarray}
\Gamma _{\phi  \to \chi \chi } = \frac{\nu ^2 \sigma ^2}{8\pi  m_\phi }&,&\Gamma _{\phi  \to \bar \psi \psi } = \frac{h^2 m_\phi }{8\pi } .
\label{eq38}
\end{eqnarray}
 Considering $ \nu\sigma $ and $ h^2 $ small for simplicity, we can obtain $ \Gamma _{tot} = \Gamma(\phi  \to \chi \chi) + \Gamma(\phi  \to \bar \psi \psi) < 3H(t) $.
In this process the energy density of the field $ \phi $ decreases (due to expansion of the Universe), therefore at the time $ t^* $ , $ 3H(t^*) $ becomes less than $ \Gamma _{tot} $. So the contribution of produced particles to the total matter density becomes significant at the time $ t^* $ and the condition $ {3H(t^*) \sim \Gamma _{tot}} $ leads to the following condition on $ \rho ^* = \rho(t^*) $, the energy density at time $ t^* $,
\begin{equation}
 \rho ^* \sim \Gamma _{tot} M_p \sqrt {\frac{\lambda }{12\pi }} .
 \label{eq39}
\end{equation}
On the other hand, considering strong enough interaction between $ \chi $ and $ \psi $-particles, after thermodynamical equilibrium, one can obtain temperature $ T_R $
\begin{equation}
 \rho ^* \sim \frac{\pi ^2 N(T_R)}{30} T_R^4 ,
 \label{eq40}
\end{equation}

where $N(T_R)$ is the effective number of degrees of freedom at $T=T_R$, with $ N(T_R) \sim 10^2-10^3$.\\
 So the reheating temperature for our model, F-term potential on $\textit{RS II}$ brane is:
\begin{equation}
T_R \sim \left( {\frac{15\Gamma M_p^2}{\pi ^2N(T_R)} \sqrt {\frac{3\lambda }{\pi }} } \right)^{\frac{1}{4}} .
\label{eq41}
\end{equation}
Considering leptogenesis which means that the inflaton field $ \phi $ decays to massive right handed neutrinos $ \nu_R $ or \textit{sterile neutrinos} $ \psi $, $ \phi \longrightarrow \psi + \psi $, so the decay rate of $ \phi $ is \cite{32}
\begin{equation}
\Gamma  = \sqrt {\frac{2}{\xi}} \frac{g}{8 \pi} M_\psi ^2 ,
\label{eq42}
\end{equation}
where $ M_\psi $ is the sterile neutrino mass and we choose a favorable mass range, $ {M_\psi \sim 10^9 Gev} $, because it is accessible by high energy experiments, such as $ LHC $. At the other hand if $ M_\psi $ is near the electroweak scale, the origin of both scales may be related and the origin of matter may be explained by leptogenesis from CP-violating $ \nu_R $ oscillations \cite{33}. For a thermal leptogenesis, reheating temperature is \cite{33}
\begin{equation}
T_R \geq 2 \times 10^9 Gev .
\label{eq43}
\end{equation}
So with the values of coupling constant , $ g $, and the Fayet-Iliopoulos term, $ \xi $, used in section(\ref{sec5}), we find a lower limit on brane tension, $ \lambda $,
\begin{equation}
\lambda \geq 9.821 \times 10^{-48} M_p^4 .
\label{eq44}
\end{equation}
with this result we find a lower limit for the brane tension $ \lambda $.
\section{Preheating after Inflation}
\begin{figure}[!ht]
\centering
\includegraphics[scale=0.328]{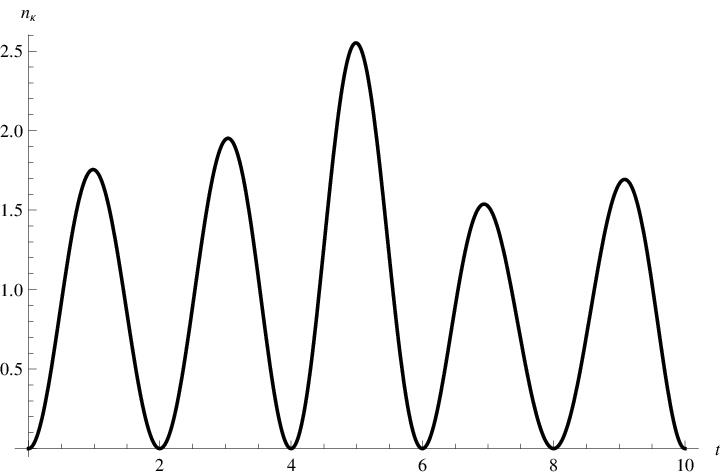}
\caption{Occupation number $ n_\kappa $ for $ (q = 0.0001 ; \kappa = 0.5) $}
\label{fig1}
\end{figure}

\begin{figure}[!ht]
\centering
\includegraphics[scale=0.328]{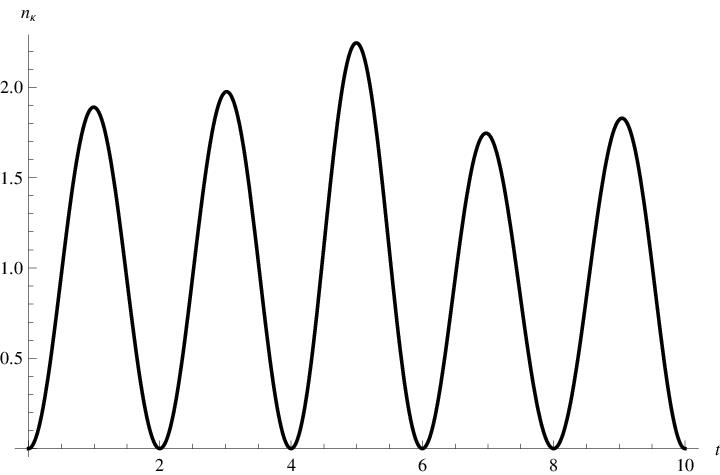}
\caption{Occupation number $ n_\kappa $ for $ (q = 0.0001 ; \kappa = 0.9) $}
\label{fig2}
\end{figure}

It is important to note that $ T_R $ is not necessarily the largest temperature reached in the thermal history of the Universe and some cases which inflaton may be kinematically forbidden to decays\cite{34}, the temperature can be much higher \cite{35}. In such cases $ T_R $ is independent of the couplings $ h,\nu $ and only depends on $ m_\phi $. Then the effective masses, after affecting quantum corrections, can have a strong effect on the dynamics of the system. Considering that effective masses can be time and space dependent, leads us to preheating \cite{31}.\\
In this work, we concentrate on \textit{Fermionic preheating} which considered the possibility of resonant production of fermions. Since many problematic particles Exp. gravitinos, are fermions and resonant production of them can have a deep impact on relic abundances \cite{31}, \cite{36}, \cite{37}, Fermionic preheating can be so important and useful. Since fermions obey the exclusion principle which means $ n_k \leq 1 $, the system is strongly constrained.\\
The major idea of fermionic preheating includes oscillation of homogeneous scalar field $ \phi $, inflaton, in an expanding flat FRW universe, which triggered creation of Dirac fermions. With appropriate matter action, one must solve the Heisenberg equation of motion for some Dirac $ \psi $ -field in cosmological background. The matter action is \cite{38}
\begin{equation}
\begin{split}
&{S_M}[\phi ,\psi ] =
\\ &
 \int {{d^4}x \left[ \frac{1}{2} \partial _\mu \phi  \partial ^\mu \phi  - V(\phi ) 
+ i\bar \psi \gamma ^\mu D_\mu \psi  - \left( m_\psi  + h\phi  \right)\bar\psi \psi  \right] } ,
\end{split}
\label{eq45}
\end{equation}
where h is the coupling constant of Yukawa coupling between $ \psi $ and $ \phi $, $ \gamma ^\mu $ is Dirac gamma matrix and $ {D_\mu } = {\partial _\mu } + \frac{1}{4}{\gamma _{\alpha \beta }}\omega _\mu ^{\alpha \beta } $ is the $ spin- \frac{1}{2} $ covariant derivative with spin connection, $ \omega _\mu ^{\alpha \beta } $ and $ {\gamma _{\alpha \beta }} \equiv {\gamma _{[\alpha }}{\gamma _{\beta ]}} $.\\
Variation of action (\ref{eq45}) with respect to $ \bar \psi $ gives us the GR generalized Dirac equation
\begin{equation}
[i{\gamma ^\mu }{D_\mu } - \left( {{m_\psi } + h\phi (t)} \right)]\psi (x) = 0 .
\label{eq46}
\end{equation}
If we consider appropriate gamma matrices for FRW space-time and decompose $ \psi(x) $ into eigen-spinors which the time dependence of the eigen-spinors is considered in some mode function $ {\chi _k}(t) $, the mode equation will be obtained
\begin{equation}
\begin{split}
&{{\ddot \chi }_k} + 4\frac{{\dot a}}{a}{{\dot \chi }_k} 
\\ &
+ \left[ {\frac{{{k^2}}}{{{a^2}}} + M_{eff}^2 - i\frac{{\left( {a{{\dot M}_{eff}}} \right)}}{a} + \frac{9}{4}{{\left( {\frac{{\dot a}}{a}} \right)}^2} + \frac{3}{2}\frac{{\ddot a}}{a}} \right]{\chi _k} = 0 ,
\end{split}
\label{eq47}
\end{equation}
where the effective mass of fermion is $ {M_{eff}} \equiv \left( {{m_\psi } + h\phi } \right) $. we can remove damping term by using conformal transformation $ {X_k}(t) = {a^2}{\chi _k}(t) $, and the mode equation reduces to
\begin{equation}
{{\ddot X}_k} + \left[ {\frac{{{k^2}}}{{{a^2}}} + M_{eff}^2 - i\frac{{{{\left( {a{M_{eff}}} \right)}^.}}}{a} + \Delta (a)} \right]{X_k} = 0 ,
\label{eq48}
\end{equation}
where $ \Delta (a) \equiv \left[ {\frac{1}{4}{{\left( {\frac{{\dot a}}{a}} \right)}^2} - \frac{1}{2}\frac{{\ddot a}}{a}} \right] $.\\
In order to calculate the number of particles created, we must use creation and annihilation operators, $ ({a_{k,s}},{b_{k,s}}) $ and $ ({a^\dag }_{k,s},{b^\dag }_{k,s}) $ . At time $ t=0 $ Hamiltonian 
\begin{equation}
H = \int {{d^3}x\left[ {i{\psi ^\dag }\dot \psi } \right]}
\label{eq49}
\end{equation}
is diagonal in terms of creation and annihilation operators, but for later times we should use Bogoliubov transformation to have a diagonal Hamiltonian in terms of creation and annihilation operators. Now we write mode function $ X_k $ in the adiabatic form with Bogoliubov coefficients $ \alpha_k $ and $ \beta_k $
\begin{equation}
{X_k}(t) = {\alpha _k}{N_ + }{e^{ - i\int\limits_0^t {dt{\Omega _k}(t)} }} + {\beta _k}{N_ - }{e^{ + i\int\limits_0^t {dt{\Omega _k} ,(t)} }}
\label{eq50}
\end{equation}
where $ {N_ \pm } = {\left( {2{\Omega _k}\left( {{\Omega _k} \pm {M_{eff}}} \right)} \right)^{ - 1/2}} $. Also we need some initial conditions to solve mode equation, (eq.\ref{eq48}), which will be obtained from normalized $ X_k $, that $ n_k(0)=0 $ and $ {X_k}({0^ - }) = {N_ + }{e^{ - i{\Omega _k}t}} $. So the positive frequency initial condition is $ {N_ + } = {\left( {2{\Omega _k}\left( {{\Omega _k} + {M_{eff}}} \right)} \right)^{ - 1/2}} $,\cite{38} . After solving mode equation (eq.\ref{eq48}) we can write the co-moving number density of created particles which determined by $ n_k(t) = |\beta_k|^2 $, and the energy density of created fermions $ \rho_\psi $ as
\begin{equation}
\begin{split}
{n_k}(t) = &
 a\left( {\frac{{{\Omega _k} - {M_{eff}}}}{{2{\Omega _k}}}} \right)\times
\\ &
 \left[ {|{{\dot X}_k}{|^2} + \Omega _k^2|{X_k}{|^2} - 2{\Omega _k}{\mathop{\rm Im}\nolimits} \left( {{X_k}\dot X_k^*} \right)} \right] ,
\end{split}
\label{eq51}
\end{equation}
\begin{equation}
{\rho _\psi }(t) = \frac{1}{{\pi {a^3}}}\int {dk{k^2}{\Omega _k}{n_k}(t)} ,
\label{eq52}
\end{equation}
where $ \Omega _k^2 \equiv \frac{{{k^2}}}{{{a^2}}} + M_{eff}^2 $.\\
We can solve equation (eq.\ref{eq48}) with considering the expansion of the universe or without it. First we attempt to find an analytical solution for this equation without expansion of the universe. For this purpose, we suppose inflaton field as an oscillating scalar field with frequency $ \omega_{\phi} $ after inflation. With $ \omega_{\phi} \gg H $ ,  the rate of cosmic expansion, we can easily neglect the expansion of the universe and time dependence of the scale factor in the mode equation (eq.\ref{eq48}) \cite{38}. So scale factor is constant and we can scale it to $ a=1 $. It is convenient to consider a simple potential like quadratic potential $ V(\phi ) = \frac{1}{2}m_\phi ^2{\phi ^2} $ and use parametric resonance method to solve this mode equation as in \cite{38}, which $ \phi $ contains an time-independent amplitude and a time-dependent part
\begin{equation}
\begin{array}{*{20}{c}}
{\phi (t) = {\phi _0}f(t),}&{f(t) = \cos \left( {{m_\phi }t} \right)} ,
\end{array}
\label{eq53}
\end{equation}
which $ m_\phi $ is constant. For more complicated potentials , like F-term model, $ m_\phi $ depends on $ \phi $ and $ t $ :
\begin{equation}
{m_\phi ^2 = \frac{{{\partial ^2}V}}{{\partial {\phi ^2}}} = {{\left( {\frac{1}{2}g\xi } \right)}^2}\left( {\frac{{ - {g^2}}}{{2{\pi ^2}{\phi ^2}}} + \frac{{3{\phi ^2}}}{{{M^4}}}} \right)}
\end{equation}
\begin{equation}
\begin{array}{*{20}{c}}
\Rightarrow &{{m_\phi } = \left( {\frac{{g\xi }}{2}} \right)}
\sqrt {\frac{{3{\phi ^2}}}{{{M^4}}} - \frac{{{g^2}}}{{2{\pi ^2}{\phi ^2}}}} .
\end{array}
\label{eq54}
\end{equation}
Since this potential exponentially grows about $ \phi=0 $, we suppose that field $ \phi $ oscillates around the minimum potential which is a complex amount,
\begin{equation}
{\phi ^4} = \frac{{ - {g^2}}}{{8{\pi ^2}}} \Rightarrow {\phi _{\min }} = \sqrt[4]{{\frac{{{g^2}}}{{8{\pi ^2}}}}}\left( {1 + i} \right) ,
\label{eq55}
\end{equation}
and we need real part of it , $ {\mathop{\rm Re}\nolimits} \left( {{\phi _{\min }}} \right) = \sqrt[4]{{\frac{{{g^2}}}{{8{\pi ^2}}}}} $ to avoid that the potential diverges. At the other hand, the oscillation is small enough to make a small variation in field $ \phi $ and mass $ m_{\phi} $ and allow us to use a periodic mean value as Effective mass $ {{\bar m}_\phi } $ in the form of
$ {{\bar m}_\phi } = \frac{1}{T}\int_{t'}^{t' + T} {dt'{m_\phi }(t')} $
which is big enough to set $ \omega_{\phi} = {{\bar m}_\phi } $. In this method effective mass of scalar field $ \phi $ must be real; so
\begin{equation}
\phi  = {\phi _0}\cos \left( {{\omega _\phi }t} \right) + {\phi _{\min }},{\phi _0} < {\phi _{\min }} .
\label{eq56}
\end{equation}
Now we define a dimensionless time variable $ \tau  \equiv{\bar m}_\phi t $ and can rewrite the mode equation (eq.\ref{eq48}) : 
\begin{equation}
{{\chi ''}_k} + \left[ {{\kappa ^2} + {{\left( {\tilde m + \sqrt q \cos \tau } \right)}^2} + i\sqrt q \sin \tau } \right]{\chi _k} = 0 ,
\label{eq57}
\end{equation}
where we have introduced the dimensionless momentum $ \kappa  \equiv \frac{k}{{{{\bar m}_\phi }}} $ ,  the dimensionless fermion mass $ \tilde m \equiv \frac{{{m_\psi }}}{{{{\bar m}_\phi }}} $ , and the resonance parameter $ q \equiv \frac{{{h^2}\phi _0^2}}{{{{\bar m}_\phi }}} $ as have introduced in \cite{38}. Attempting to solve this complex mode equation and considering some numerical results presented in \cite{38}, we split the mode function $ \chi_{k} $ into it's real and imaginary part, $ 
{\chi _k}\left( \tau  \right) = 
\sin \left( {{\nu _k}\tau } \right)\left[ {{\chi _{1k}}\left( \tau  \right) + {\rm{ }}i{\chi _{2k}}\left( \tau  \right)} \right] $ and the equation (eq.\ref{eq57}) separates to a system of coupled second-order differential equations :
\begin{equation}
\begin{split}
{{\chi ''}_{1k}} &+ \left( { - \nu _k^2 + {\kappa ^2} + {{\left( {\tilde m + \sqrt q \cos \tau } \right)}^2}} \right){\chi _{1k}} 
\\ &
- \sqrt q \sin \tau {\chi _{2k}} = 0 ,
\end{split}
\label{eq58}
\end{equation}
\begin{equation}
\begin{split}
{{\chi ''}_{2k}}& + \left( { - \nu _k^2 + {\kappa ^2} + {{\left( {\tilde m + \sqrt q \cos \tau } \right)}^2}} \right){\chi _{2k}}
\\ &
 + \sqrt q \sin \tau {\chi _{1k}} = 0 ,
\end{split}
\label{eq59}
\end{equation}
Following the calculations of Appendix A, one can solve this equations and obtain
\begin{equation}
{\chi _{1k}} = {c_1}{e^{{\lambda _1}\tau }} + {c_2}{e^{{\lambda _2}\tau }} + {c_3}{e^{{\lambda _3}\tau }} + {c_4}{e^{{\lambda _4}\tau }} ,
\label{eq60}
\end{equation}
\begin{equation}
{\chi _{2k}} = i\left[ { - {c_1}{e^{{\lambda _1}\tau }} - {c_2}{e^{{\lambda _2}\tau }} + {c_3}{e^{{\lambda _3}\tau }} + {c_4}{e^{{\lambda _4}\tau }}} \right] ,
\label{eq61}
\end{equation}
where
$ {\lambda _{1,2}} =  \pm \sqrt {{\mathop{\rm R}\nolimits} \left( {{e^{i{\theta _ + }}}} \right)} $ , 
$ {\theta _ + } \equiv arctan\left( {\frac{{{l_2}}}{{{l_1}}}} \right) $ ,
$ {l_1} \equiv \nu _k^2 - {\kappa ^2} - {\left( {\tilde m + \sqrt q \cos \tau } \right)^2} $,
$ {l_2} \equiv - \sqrt q \sin \tau $ and 
$ R \equiv \sqrt {l_1^2 + l_2^2} $ . So
\begin{equation}
{\chi _k} = \sin \left( {{\nu _k}\tau } \right)\left( {2{c_1}{e^{{\lambda _1}\tau }} + 2{c_2}{e^{{\lambda _2}\tau }}} \right) .
\label{eq62}
\end{equation}
With inserting 
$ \left| {\dot \chi _k^2} \right| $,
$ \left| {\chi _k^2} \right| $ and 
$ {\mathop{\rm Im}\nolimits} \left( {{\chi _k}\dot \chi _k^*} \right) $
in (eq.\ref{eq51}) one can obtain $ n_{k}(t) $ in terms of $ c_{1}^{2} $, $ c_{2}^{2} $ and $ c_{1}c_{2} $.
Using positive frequency initial conditions to normalize mode function $ \chi_{\kappa} $ results :
\begin{widetext}
\[
{c_1} = \frac{1}{{4{{\bar m}_\phi }\sqrt {2\left( {{\kappa ^2} + {{\tilde m}^2} + 2\sqrt q  + q + \left( {\tilde m + \sqrt q } \right)\sqrt {{\kappa ^2} + {{\tilde m}^2} + 2\sqrt q  + q} } \right)} }} , \]
\end{widetext}
\begin{equation}
{c_1} =  - {c_2} ,
\label{eq63}
\end{equation}
and
\begin{equation}
{\chi _\kappa} = 2{c_1}\sin \left( {{\nu _\kappa}\tau } \right)\left( {{e^{{\lambda _1}\tau }} - {e^{ - {\lambda _1}\tau }}} \right) .
\label{eq64}
\end{equation}
To determining the frequency $ \nu _k $ , we consider the value of mode function $ \chi _k ( T ) = \cos d_k + i \sin d_k $ exactly after the first background oscillation :
\begin{equation}
\cos {d_\kappa} =  - \cos {\nu _\kappa}T = 0 \Rightarrow {\nu _\kappa}T = \frac{\pi }{2} ,
\label{eq65}
\end{equation}
and we scale period $ T $ to 1; so we have $ {\nu _\kappa} = \frac{\pi }{2} $ .\\
As you can see in Fig.\ref{fig1} , Fig.\ref{fig2} , for small enough amounts of $ q $, the occupation number $ n_{\kappa} $ often exhibits smooth behavior. Analytical solution we find in this paper , have a good compatibility with previous numeric results  for  $ q \le 0.001 $ , as in \cite{38} .
\section{Conclusion and Discussion}
In this paper, we have analyzed F-term hybrid inflation in the framework of $ RS II $ braneworld and one can easily show that this setup can provide successful inflation. From the point of view of the reheating, we compute the reheating temperature, $ T_R $, for thermal leptogenesis in this model. Using obtained results, we find some constraints on brane tension, $ \lambda $. At the other hand, we show some inflation perturbation parameters i.e. $ n_s $, $ r $, $ dn_s/d \ln k $ have good compatibility with $ \textit{WMAP 9-year} $ and $ Planck $ observations.\\
Since we are studying thermal history of the universe, it is so important to have a suitable end of inflation and entering the matter production era. We looked for an appropriate preheating mechanism for this setup and computed fermionic preheating to arising an efficient leptogenesis. Analytical solution we find in this paper, have a good compatibility with previous numeric results, \cite{38}.
\appendix
\section{second order systems}
As mentioned in \cite{39}, Given a second order $ n n $ system $ {x''} = Ax $ , define the variable $ u $ and
the $ 2n \times 2n $ block matrix $ C $ as follows.
\begin{equation}
u = \left[ {\begin{array}{*{20}{c}}
x\\
{x'}
\end{array}} \right] , C = \left[ {\begin{array}{*{20}{c}}
0&I\\
A&0
\end{array}} \right] .
\label{eq66}
\end{equation}
Then each solution $ x $ of the second order system $ {x''} = Ax $ produces a corresponding solution $ u $ of the first order system $ {u'} = Cu $. Similarly, each solution $ u $ of $ {u'} = Cu $ gives a solution $ x $ of $ {x''} = Ax $ by the formula $ x = diag\left( {I,0} \right)u $.\\
The characteristic equation for this system is
\begin{equation}
\det \left( {C - \lambda I} \right) = {\left( { - 1} \right)^n}\det \left( {A - {\lambda ^2}I} \right) .
\label{eq67}
\end{equation}
After solving characteristic equation
$ \det \left( {C - \lambda I} \right) = 0 $
and calculating eigenvalues enable us to finding eigenvectors of $ A(n \times n) $ and $ C(2n \times 2n) $, which is
\begin{equation}
\left( {C - \lambda I} \right)\left( {\begin{array}{*{20}{c}}
W\\
Z
\end{array}} \right) = 0 \Leftrightarrow \left\{ {\begin{array}{*{20}{c}}
{AW = {\lambda ^2}W}\\
{Z = \lambda W}
\end{array}} \right .
\label{eq68}
\end{equation}
Considering the eigen-pairs of matrix $ C $ , 
$ \left\{ {\left( {{\lambda _j},{y_j}} \right)} \right\}_{j = 1}^{2n} $ and
$ {y_1}, \ldots ,{y_{2n}} $
are independent, one can obtain general solutions of $ u′ = Cu $ and $ x′′ = Ax $. Introducing $ {w_j} = diag\left( {I,0} \right){y_j} $ and 
$ j = 1, \ldots ,2n $, we have :
\begin{equation}
\begin{array}{l}
\begin{array}{*{20}{c}}
{u(t) = {c_1}{e^{{\lambda _1}\tau }}{y_1} +  \ldots  + {c_{2n}}{e^{{\lambda _{2n}}\tau }}{y_{2n}}}&{(2n \times 1) ,}
\end{array}\\
\begin{array}{*{20}{c}}
{x(t) = {c_1}{e^{{\lambda _1}\tau }}{w_1} +  \ldots  + {c_{2n}}{e^{{\lambda _{2n}}\tau }}{w_{2n}}}&{(n \times 1) .}
\end{array}
\end{array}
\label{eq69}
\end{equation}
In our case
\begin{equation}
\left[ {\begin{array}{*{20}{c}}
{{{\chi ''}_{1k}}}\\
{{{\chi ''}_{2k}}}
\end{array}} \right] = \left[ {\begin{array}{*{20}{c}}
{l_1 }&{- l_2 }\\
{l_2 }&{ l_1}
\end{array}} \right]\left[ {\begin{array}{*{20}{c}}
{{\chi _{1k}}}\\
{{\chi _{2k}}}
\end{array}} \right] .
\label{eq70}
\end{equation}
We introduce 
$ {l_1} \equiv  \nu _k^2 - {\kappa ^2} + {\left( {\tilde m + \sqrt q \cos \tau } \right)^2} $ and 
$ {l_2} \equiv - \sqrt q \sin \tau $, 
then we have
\begin{equation}
C = \left[ {\begin{array}{*{20}{c}}
0&0&1&0\\
0&0&0&1\\
{{l_1}}&{ - {l_2}}&0&0\\
{{l_2}}&{{l_1}}&0&0
\end{array}} \right] .
\label{eq71}
\end{equation}
The result is obtained by solving characteristic equation 
\begin{equation}
\det \left( A - {\lambda ^2}I \right) = 0 \Rightarrow \left\{ {\begin{array}{*{20}{c}}
{\lambda _{1,2} =  \pm \sqrt {{l_1} + i{l_2}} }\\
{\lambda _{3,4} =  \pm \sqrt {{l_1} - i{l_2}} } 
\end{array}} , \right.
\label{eq72}
\end{equation}
and we can rewrite them in the form of
\begin{equation}
\begin{array}{*{20}{c}}
{{\lambda _{1,2}} =  \pm \sqrt {{\mathop{\rm R}\nolimits} \left( {{e^{i{\theta _ + }}}} \right)} }&{,}&{{\lambda _{3,4}} =  \pm \sqrt {{\mathop{\rm R}\nolimits} \left( {{e^{i{\theta _ - }}}} \right)} }&{,}
\end{array}
\label{eq73}
\end{equation}
where 
$ R \equiv \sqrt {l_1^2 + l_2^2} $,
$ {\theta _ + } \equiv arctan\left( {\frac{{{l_2}}}{{{l_1}}}} \right) $ and
$ {\theta _ - } \equiv arctan\left( {\frac{{ - {l_2}}}{{{l_1}}}} \right) $ .
Following the method we can obtain eigenvectors and $ w_{j} $, and finally we have
\begin{equation}
\begin{array}{l}
{\chi _{1k}} = {c_1}{e^{{\lambda _1}\tau }} + {c_2}{e^{{\lambda _2}\tau }} + {c_3}{e^{{\lambda _3}\tau }} + {c_4}{e^{{\lambda _4}\tau } , }\\
{\chi _{2k}} = i\left[ { - {c_1}{e^{{\lambda _1}\tau }} - {c_2}{e^{{\lambda _2}\tau }} + {c_3}{e^{{\lambda _3}\tau }} + {c_4}{e^{{\lambda _4}\tau }}} \right] .
\end{array}
\label{eq74}
\end{equation}
One may concern whether $ l_{1} $ and $ l_{2} $, Coefficients of the equations , become zero or not and how affect the form of equation and solution. So we find zero points of them and obtain special meaningful solutions in our case ; if $ l_{1}=0 $ :
\begin{equation}
\begin{split}
{l_1} = 0 & \Rightarrow {l_2} =  - \sqrt {q - q{{\left( {\tilde m \mp \sqrt {\nu _k^2 - {\kappa ^2}} } \right)}^2}} 
\\ &
 \Rightarrow \left\{ {\begin{array}{*{20}{c}}
{{\theta _ + } = arctan\infty }\\
{{e^{i{\theta _ + }}} =  \pm i} 
\end{array}} , \right.
\end{split}
\label{eq75}
\end{equation}
and the mode function is 
\begin{equation}
{\chi _k}\left( \tau  \right) = \sin \left( {{\nu _k}\tau } \right)\left( {2{c_1}{e^{ \pm i\sqrt {{l_2}} \tau }} + 2{c_2}{e^{ \mp i\sqrt {{l_2}} \tau }}} \right) .
\label{eq76}
\end{equation}
And if $ l_{2}=0 $ :
\begin{equation}
\begin{split}
{l_2} = 0 & \Rightarrow {l_1} = \nu _k^2 - \left( {{\kappa ^2} + {{\left( {\tilde m + \sqrt q \cos n\pi } \right)}^2}} \right)
\\ &
 \Rightarrow \left\{ {\begin{array}{*{20}{c}}
{{\theta _ + } = arctan0}\\
{\cos {\theta _ + } = \cos n\pi }
\end{array}} , \right.
\end{split}
\label{eq77}
\end{equation}
with mode function in the form of
\begin{equation}
{\chi _k}\left( \tau  \right) = \sin \left( {{\nu _k}\tau } \right)\left( {2{c_1}{e^{i\sqrt {{l_1}} \tau \cos n\pi }} + 2{c_2}{e^{i\sqrt {{l_1}} \tau \cos n\pi }}} \right) .
\label{eq78}
\end{equation}

\end{document}